\documentclass[aps,showpacs,floats,epsfig]{revtex4}
\usepackage{epsfig}
\usepackage{amsmath}
\usepackage{amsfonts}
\usepackage{graphicx}
\usepackage{amssymb}
\usepackage{amsbsy}
\usepackage{xcolor}
\usepackage{subfigure}
\usepackage{setspace}% ,caption}
\usepackage{hyperref}
\usepackage[autostyle]{csquotes}
\usepackage[up,bf,raggedright]{titlesec}

\newcommand{\ket}[1]{|{#1}\rangle}
\newcommand{\braket}[1]{\langle{#1}\rangle}

\makeatletter
\gdef\@ptsize{0} 
\makeatother
\begin{document}

\title {A single atom noise probe operating beyond the Heisenberg limit}

\author{T.~Dutta}
\email[Electronic address: ]{cqttaru@nus.edu.sg}
\affiliation{Centre for Quantum Technologies, National University
Singapore, Singapore 117543}
\author{M.~Mukherjee}
\email[Electronic address: ]{phymukhe@nus.edu.sg}
\affiliation{Centre for Quantum Technologies, National University
Singapore, Singapore 117543} \affiliation{Department of Physics,
National University Singapore, Singapore 117551. }
%
%\date{\today}
\maketitle

\pagebreak

\doublespacing
\section{Abstract}
%\begin{abstract}
According to the Heisenberg uncertainty principle, the energy or frequency uncertainty of a measurement can be at the best inversely proportional to the observation time~($T$). The observation time in an experiment using a quantum mechanical probe is ultimately limited by the coherence time of the probe. Therefore the inverse proportionality of the statistical uncertainty of a frequency measurement to the observation time is also limited up to the coherence time of the probe, provided the systematic uncertainties are well below the statistical uncertainties. With a single laser-cooled barium ion as a quantum probe, we show that the uncertainty in the frequency measurement for a general time-dependent Hamiltonian scales as $1/T^{1.75\pm 0.03}$ as opposed to $1/T$, given by the Heisenberg limit for time-independent Hamiltonian. These measurements, based on controlled feedback Hamiltonian and implemented on a laser cooled single ion, allowed precise measurement of noise frequency in the kHz range. Moreover, based on the observed sensitivity of a single ion experiment presented here, we propose the use of a similar protocol with enhanced sensitivity as a tool to directly verify the existence of certain types of light mass axion-like dark matter particles where no direct measurement protocol exists. 

%\end{abstract}

%\pacs{32.70.Cs, 37.10.Ty, 06.30.Ft}

%
\section{Introduction}
Among all the fundamental quantities frequency or time has been measured with the smallest uncertainty ~\cite{Brewer2019}. The most precise estimate of frequency is obtained by comparing the phase evolution of a nearly single-frequency laser to that of an isolated two-level atom prepared in a superposition state. The rate of change of the acquired phase difference (also known as a phase estimator) between the two systems provides an estimate of their frequency difference. Thus the laser frequency can be compared to the excitation frequency of a two-level atom, assuming that the energy difference between the two levels of the atom is not changing during the measurement time. As the phase accumulation for such an ideal system is proportional to the integration time, the precision with which the frequency difference can be estimated is inversely proportional to the integration time. However, in the case of real systems, one needs to consider rigorous error analysis as detailed below.

In error analysis, the total uncertainty or the precision of any parameter estimation is limited by the uncertainty of the estimator that is used to estimate the parameter~\cite{Helstrom1976,Holevo1982}. This uncertainty originates from: (i) the inability to control the influence of external disturbances on to the probe and (ii) the statistical distribution of the experimental outcome by repetitive measurement of the estimator. The former, also known as the systematic uncertainty, is dependent on the experimental setup whereas the latter, called the statistical uncertainty is more fundamental and is of our concern here. The statistical uncertainty can be reduced by averaging over many measurements and by increasing the observation time for each measurement. The dependence of the statistical uncertainty on the observation time has a fundamental origin related to the Heisenberg energy-time uncertainty principle.
 
In this work, we experimentally show that it is possible to measure frequency with a precision that scales faster the inverse linear scaling with observation time~\cite{Pang2017}. The proposed method is based on coherent control feedback to acquire phase information at a faster rate. However, it is not applicable to an atomic clock where the frequency is time-independent. Therefore our protocol's applicability lies in noise frequency estimation in a clock transition with quadratically higher precision as compared to any known techniques for the same observation time. This motivates us to apply this new protocol to search for exotic particles that interact with an atomic electron in a time-dependent manner, much like a noise term acting on a clock transition.

In the remaining part of the introduction, we will first discuss the theoretical bounds on frequency estimation followed by a discussion on different approaches to overcome such theoretical bounds.\\
%%%%%%%% Figure 1
%%%%%%%
%
%\subsection*{Theoretical bounds on frequency estimation}
%
A parameter, e.g. frequency can be estimated in an experiment that involves three stages of experimentation~\cite{Giovannetti2011}: (a) preparation of the probe (b) interaction of the probe with the environment and (c) probe readout. By repeating these three stages, the statistical uncertainty can be reduced upto the asymptotic limit given by the central-limit theorem which states that if $\sigma$ is the standard deviation of a single measurement, averaging over a large number of independent measurements ($n$), the standard deviation converges to $\sigma/\sqrt{n}$. Adopting different strategies such as use of entangled probes~\cite{Wineland1992,Ma2009,Napolitano2011} instead of independent probes, it is possible to surpass the {\it Shot Noise Limit}~(SNL). However, for a single probe, quantum mechanics provides an ultimate lower bound on the uncertainty of an estimator called the Heisenberg bound. For a time-independent Hamiltonian, the Heisenberg bound can be achieved by adopting proper quantum strategies as shown in ref.~\cite{Giovannetti2011} and \cite{Giovannetti2004}. From the information theoretic point of view, the bound on the mean-squared deviation of an unbiased estimator $\hat{\alpha}$ is given by~\cite{Helstrom1976,Holevo1982,Fisher1925}
\begin{equation}
\braket{\delta^2\hat{\alpha}}\geq 1/nF_{\alpha}, \label{eq:var}
\end{equation}          
     
where, $n$ is asymptotically large number of independent measurements while $F_{\alpha}$ is the Fisher information associated with the estimator for the parameter $\alpha$. The Fisher information is a measure of the information content in the measured data about the parameter $\alpha$ and the bound is known as the Cram\'er-Rao bound~(CRB)~\cite{Cramer1946}. In case the probe is prepared in a superposition state $\ket{\psi_{\alpha}}$, the corresponding quantum Fisher information~(QFI) is given~\cite{samuel1994,samuel1996} by
\begin{equation}
F_{\alpha} = 4(\braket{\partial_{\alpha}\psi|\partial_{\alpha}\psi}-|\braket{\psi|\partial_{\alpha}\psi}|^2), \label{eq:QFI}
\end{equation}   
which maximizes the classical Fisher information over all the possible types of quantum positive operator valued measurements~(POVM) on the state. This implies that for a parameter $\alpha$ which is a multiplicative factor to a time-independent Hamiltonian and whose time evolution is given by an unitary as $U_{\alpha}=\exp(-i\alpha Ht)$, the QFI $F_{\alpha}$ as obtained from eq.(\ref{eq:QFI}) is $\propto T^2$ for a phase evolution time (also called the observation time) $T$. Thus from eq.~(\ref{eq:var}) one obtains an inverse linear scaling of the uncertainty (defined as the square-root of the variance) on parameter $\alpha$ with observation time. In the case of an atomic clock, the parameter is frequency which is best estimated using the Ramsey measurement scheme leading to a precision that scales as $1/T$.//
%
%\subsubsection*{Alternative approaches to overcome theoretical bounds on frequency estimation}
%
We have discussed how the shot noise limit can be surpassed by the use of correlated multiple probes. In the following, we restrict ourselves to a single probe. One way to surpass the Heisenberg limit with a single probe is to exploit the non-linear response of the Hamiltonian, as an example for a Hamiltonian $H \propto e^t\sigma_z$ the energy rises exponentially with time $t$ in a trivial way. A volume of theoretical and experimental work exists towards outperforming the Heisenberg limit using the non-linear response ~\cite{ Napolitano2011,Luis2004,Boixo2007,Roy2008,Boi2008,Pezze2009,Hall2012,Zwierz2014}. Such proposals have been recently realized with a trapped ion probe where coherence of the probe responds linearly or non-linearly on the noise power spectral density~\cite{shlomi2011}. In the non-linear regime, a frequency estimation protocol leads to a precision that is below the bound given by the Heisenberg limit ~\cite{shlomi2013}. However, in the strict sense of the Heisenberg limit it holds true for a linear response only. Pang and Yang proposed an adaptive protocol to estimate an external noise frequency component with uncertainty beyond the Heisenberg limit for a time-dependent Hamiltonian valid even in linear response regime~\cite{Pang2017,Yang2017}. This adaptive feedback approach not only surpasses the Heisenberg limit but also provides a new fundamental limit for the precision that can be achieved with a time-dependent Hamiltonian. Adopting quantum control and classical Fourier transform, it has been shown that the uncertainty scales as $1/T^{3/2}$~\cite{Schmitt2017,Boss2017} in an experiment with the single NV center. In a superconducting qubit, an uncertainty scaling of $1/T^2$ has been recently claimed~\cite{Naghiloo2017} leveraged on adaptive quantum control. In this experiment, only noise frequencies above $1~$MHz could be accessed due to the limited coherence time. A significant component of electronic noise is $1/f$-noise, where $f$ is the noise frequency component and it dominates below one MHz frequency. This regime is inaccessible to a circuit based sensor. Unlike superconducting circuits, a single atom as a probe naturally couples to different environmental perturbations with high sensitivity.
\section{Results}
In our experiment with a single ion as a quantum probe and implementing quantum feedback strategy, we demonstrate: (a) a frequency uncertainty scaling beyond the Heisenberg limit with a single atom independent of the strength of the noise; (b) two orders of magnitude improvement in probe time and hence the accessible frequency range stretches to kHz; (c) a weak radiative time-dependent coupling of the probe to the environment thus extending the sensing capability to EM fluctuations in the vicinity of the single atom probe; (d) a proof-of-principle experiment to provide a bound on the coupling strength of a low mass ($50$~kHz) axion-like particle to the atomic electron by direct measurement. To the best of our knowledge, neither a direct measurement nor a credible scheme exists to probe the existence of such a particle even though it is considered to be a valid dark matter candidate~~\cite{surjeet2013}.

Here, we introduce the theoretical framework for the quantum feedback strategy to overcome the Heisenberg bound, followed by numerical analysis and the presentation of the experimental data.  
\subsection*{Theoretical framework}
The Quantum Fisher Information~(QFI) as given in eq.(\ref{eq:QFI}) for a pure state $\ket{\psi_{\alpha}} = \sqrt{P_0}\ket{0}+e^{i\phi_{\alpha}}\sqrt{P_1}\ket{1}$ where $P_{0,1}$ are independent of $\alpha$ and $\ket{0}$, $\ket{1}$ are eigenkets of the $\sigma_z$ operator takes the form~\cite{Bures1969}
\begin{equation}
F_{\alpha} = 4P_0P_1\Big(\frac{\partial\phi_{\alpha}}{\partial\alpha}\Big)^2.\label{eq:brus}
\end{equation} 

It is important to note: (a) the QFI maximizes when the qubit state is prepared in an equal superposition state of its eigenkets (P$_0$=P$_1$=1/2) and (b) the QFI degrades if decoherence sets in. The optimal projectors as shown in~\cite{Yang2017} for maximizing the QFI in the $\sigma_z$ basis turn out to be 
\begin{equation}
\ket{\pm}_{\alpha} = \frac{1}{\sqrt{2}}(\ket{0}\pm ie^{i\phi_{\alpha_c}}\ket{1}),\label{eq:optstate}
\end{equation}

where $\alpha_c$ is the optimal value of the parameter, which indicates that a prior guess of $\alpha$ is essential in an adaptive measurement setting. This is not a predicament to the adaptive measurement scheme as a prior knowledge is gained from an estimate without the control applied. Physically the QFI as given in eq.(\ref{eq:brus}), is the sensitivity of the measured phase $\phi_{\alpha}$ as a function of the change in $\alpha$, the parameter to be estimated. This is related to the Bures distance~\cite{Bures1969} and according to eq.~(\ref{eq:var}) it is the inverse of the uncertainty of the estimator. The adaptive measurement gives desired advantage only when the parameter to be estimated is close to the actual value such that~\cite{Pang2017}
\begin{equation}
|\phi_{\alpha}-\phi_{\alpha_c}|\leq \pi/2.
\end{equation} 

In the experiment, the probe is prepared in a state $\ket{\psi_{\alpha}}$ (preparation step in the Figure~\ref{fig:theoScheme}(a)  inset (i)) and allowed to evolve it under the external time-dependent field for a certain time $T$. The final state is projected to the $\sigma_z$ basis to estimate the phase gain $\phi_\alpha$ (measurement step in the Figure~\ref{fig:theoScheme}(a) inset (i)). This is same as that of a Ramsey phase measurement sequence except that during the waiting time a time-dependent field as shown in Figure~\ref{fig:theoScheme}(a)(  inset (ii)) interacts with the atom. We intend to estimate the frequency components and corresponding amplitudes of this field. To maximize the acquired phase one applies an extra time-dependent field called the control Hamiltonian. In the following we numerically analysed the effect of such a field on our system. 
%
%%%%%%%% Figure 2
%%%%%%%%
%
\subsection*{Numerical analysis}
In ref.\cite{Yang2017,Yang_supp2017}, it has been shown that the QFI as in eq.(\ref{eq:brus}), can be boosted for a general time-dependent Hamiltonian by applying appropriate quantum control. The QFI $F_{\alpha c}$ for a general time-dependent Hamiltonian is bounded by~\cite{Pang2017}
\begin{equation}
F_{\alpha c} \le [\int_{t_0}^t (\mu_{max}(t')-\mu_{min}(t'))dt']^2, \label{eq:qfi_sat}
\end{equation} 

where $t_0$ is the initial time and $\mu_{max,min}$ refers to the maximum and minimum eigenvalues of the operator $\partial_{\alpha}H_0$ as given in eq.(\ref{eq:brus}). The equality is reached provided the initial state is prepared as an equal superposition of the eigenstates of $\partial_{\alpha}H_0$ which are $\ket{\mu_{max,min}}$ and in addition the optimal coherent control Hamiltonian is applied. 

\begin{figure*}[ht] 
 \centering
 \includegraphics[width=\textwidth]{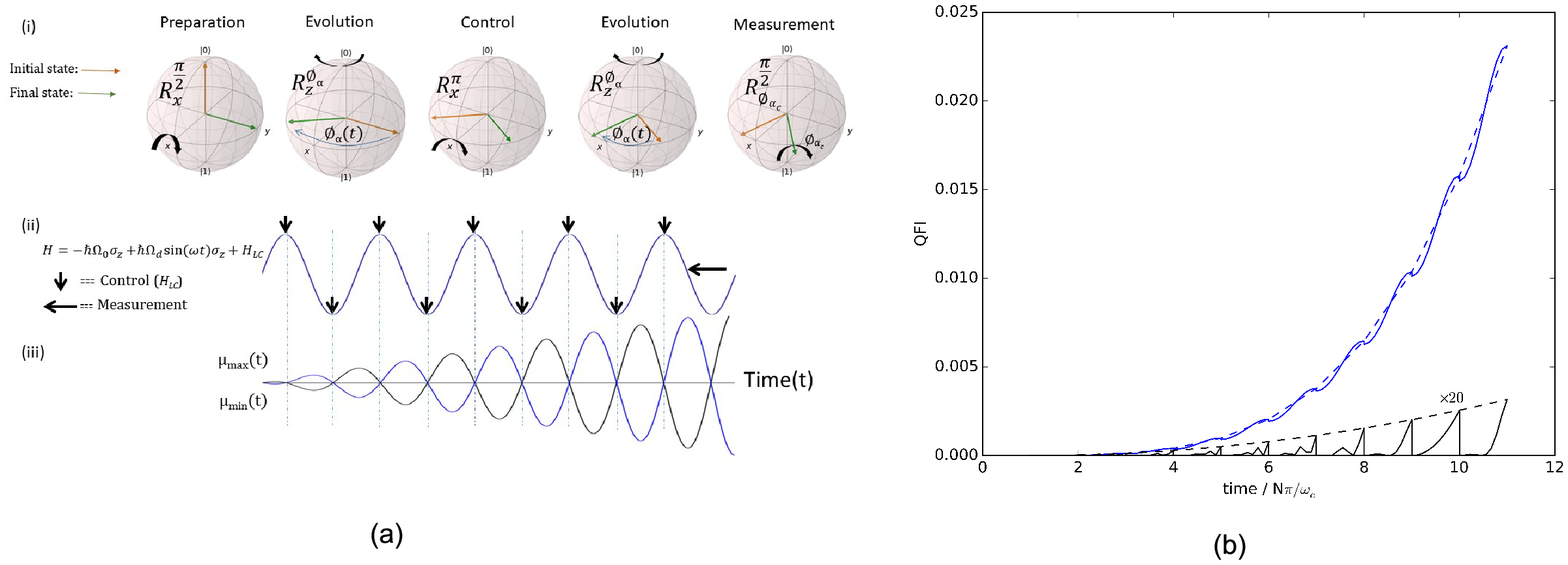} 
\caption{(a)The frequency measurement scheme using local control Hamiltonian shown in a nutshell: (i) the five steps of the experiment to estimate the frequency $\omega$ of an unknown external time-dependent field coupled to electron of an atom is shown as a qubit rotations on the Bloch sphere. The orange arrows denote the initial states while the green one denote the final state in an ideal condition; $\phi_\alpha (t)$ is the evolution of phase for time t while $\phi_{\alpha_c}$ is the optimal phase at which measurement is done (eq.~\ref{eq:optstate}).  (ii) shows the time dependence of the Hamiltonian ( eq.(\ref{eq:hamil})) where quantum control Hamiltonian (given in eq.(\ref{eq:olch})) is applied at instances marked by the vertical arrows while the optimal measurement is performed at the instant marked by the horizontal arrow; (iii) the time dependence of the eigenvalues ($\mu_{max,min}$) corresponding to eigenstates $\ket{\partial_{\omega}H}$ are shown in black and blue for $+$ and $-$  eigenvalues respectively as given in eq. (\ref{eq:freqsense}). The difference between these two eigenvalues at any time corresponds to the sensitivity of the phase change with respect to the frequency change. This shows that the control Hamiltonian always adds up the phase thus improving the sensitivity. (b) Numerically obtained Quantum Fisher Information for frequency estimation for the same Hamiltonian when no optimal control applied (black) and when both the control Hamiltonian (H$_{LC}$) and appropriate measurement protocol are applied optimally (blue). The dashed line shows the case for optimal measurements when measurement time is multiple of the frequency keeping the phase constant.}  
 \label{fig:theoScheme}  
\end{figure*}

Thus the optimal scheme for parameter estimation shown in Figure~\ref{fig:theoScheme}(a) inset (i) consists of preparation, evolution, control and measurement. In particular for the Hamiltonian as in eq.(\ref{eq:hamil}), the instantaneous maximum and minimum eigenvalues of the phase sensitivity operators ($\partial_{\omega,\Omega_d}H_0$) with respect to the frequency and amplitude estimation are:
\begin{equation}
\mu_{max,min} = \pm \hbar \Omega_d t \cos \omega t \label{eq:freqsense}
\end{equation}
and
\begin{equation}
\mu_{max,min} = \mp \hbar \sin \omega t,\label{eq:ampsense}
\end{equation}

respectively. The time evolution of these eigenvalues are shown in Figure~\ref{fig:theoScheme}(a) inset (iii) for frequency estimation. The two eigenvalues cross each other at times when the field amplitude is maximum or minimum. Thus integrating the difference of the two instantaneous eigenvalues over two consecutive cycles nearly cancel each other. In case of the amplitude estimation, the eigenvalues are shifted by $\pi/2$ phase compared to the frequency estimation. Due to the periodic crossing of the eigenvalues, the QFI as obtained by numerical integration of eq.~({\ref{eq:qfi_sat}}) leads to a slow rise ($\propto T^2$) with time as shown in Figure~\ref{fig:theoScheme}(b) black dash-line. The phase sensitivity or the QFI however can be maximized by applying an optimal level crossing Hamiltonian (OLCH) of the form~\cite{Yang2017}
\begin{equation}
H_{LC} = h(t)\sigma_x^{\pi} = \Sigma_{n=1}^N \delta(t-\frac{(2n+1)\pi}{\omega_c}),\label{eq:olch}
\end{equation}

where $n=0,1,\cdots N$ and $\omega_c$ denotes the control frequency which is optimal when $\omega=\omega_c$ and $h(t)$ is a function of time. The application of the control Hamiltonian is optimal if it is applied at the time instances shown by vertical arrows in Figure~\ref{fig:theoScheme}(a) inset (ii) which corresponds to instances where the eigenvalues cross as in Figure~\ref{fig:theoScheme}(a) inset (iii). Therefore, for frequency and amplitude estimations one needs to apply $\sigma_x^{\pi}$ pulses at the time instances when the differences in instantaneous eigenvalues are zero, thereby maximizing the QFI ($\propto T^4$) as shown in Figure~\ref{fig:theoScheme}(b) by the blue dash-line. Note that the QFI under optimal control scales as $T^4$ for frequency estimation while $T^2$ for amplitude estimation (not shown in the figure). Thus from the Figure~\ref{fig:theoScheme}(b), it is clear that without the control, the QFI scales as per the Heisenberg bound. In the case of amplitude estimation, if no control is applied the QFI is independent of the time of observation as is evident from the eq.(\ref{eq:ampsense}) but once the control is applied it scales as $T^2$.  
%%%%%%%% Figure 3
%%%%%%%%
%
\subsection*{Experimental data}
In practice, to surpass the Heisenberg limit, the systematic uncertainties are required to be below the statistical uncertainty for a given time of observation. We obtain the shot noise limited phase estimation for each experiment for any observation time that is within the qubit de-phasing time. These measurements are discussed in the Supplementary Information. The de-phasing of our qubit comes predominantly from external magnetic field noise originating from the AC power line at $50~$Hz, however the qubit has a long T$1$ time of about half of a minute. Locking the phase of our experimental cycle to the AC power line phase we obtain a de-phasing time of $500\mu$s, however this leads to long experimental time sequences. Therefore for the current experiment, the experimental cycle is not phase locked to line frequency and we obtain a de-phasing time of $\approx 80\mu$s. 

\begin{figure*}[ht] 
\centering
\includegraphics[width=\textwidth]{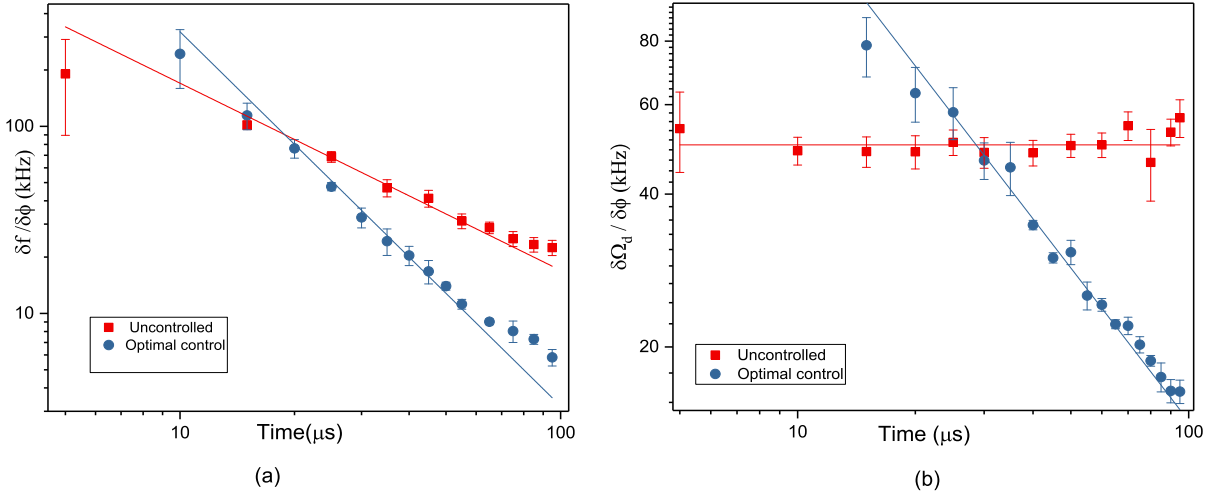} 
\caption{(a)The inverse of the frequency sensitivity ($\frac{\partial f}{\partial\phi}$; $f=\frac{\omega}{2\pi}$) measured as a function of the observation time for an externally applied time-dependent Hamiltonian of frequency $50~$kHz. The $\color{red}\blacksquare$ correspond to un-controlled measurements while the $\color{blue}\bullet$ are controlled measurements. Along with the experimental data are shown the $T^{-1}$($\color{red}\textendash$) and $T^{-2}$ ($\color{blue}\textendash$) curves as they represent the theoretical bound for un-controlled and controlled measurements respectively. The decoherence in this experiment starts playing a role around $80~\mu$s. (b) The inverse of the amplitude sensitivity measured as a function of the observation time for an externally applied time-dependent Hamiltonian of frequency $50~$kHz. The $\color{red}\blacksquare$ correspond to un-controlled measurements while the $\color{blue}\bullet$ are controlled measurements. As expected, the un-controlled measurement is independent of time of observation while the controlled one depends on observation time as $T^{-1}$ in an ideal case.} 
\label{fig:freq_sen}
\end{figure*}

The QFI as mentioned in eq.(\ref{eq:brus}) is a function of the phase sensitivity with respect to the parameter $\alpha$. Therefore instead of QFI, we measured the sensitivity with and without the OLCH as in eq.(\ref{eq:olch}) with respect to the frequency, $\frac{1}{2\pi}\frac{\partial\omega}{\partial\phi}=\frac{\partial f}{\partial\phi}$ and amplitude, $\frac{\partial\Omega_d}{\partial\phi}$. The sensitivity is obtained by measuring the change in phase as a function of the frequency or the amplitude of the external field (see the text in the Supplementary Information for further details). The measured inverse sensitivities that correspond to the variance are plotted as a function of the total observation time $T$ in Figure~\ref{fig:freq_sen}(a) for frequency and in Figure~\ref{fig:freq_sen}(b) for amplitude. In Figure~\ref{fig:freq_sen}(a), the  $\color{red}\blacksquare$ data-points ( plotted in a log-log scale) denote the measured sensitivity of the frequency measurement without applying the control Hamiltonian $H_{LC}$ for a fixed observation time. Each of these points corresponds to a sensitivity measurement as explained in the Supplementary Information. The errorbars are measured from the sensitivity data, repeated over $22$ experiments. At very short times, the errorbars are large since the phase accumulation time is short. As a comparison, $\color{red}\textendash$ line denotes the theoretical bound $1/T$. On the contrary, the $\color{blue}\bullet$ data-points correspond to the sensitivity measurements with the control Hamiltonian applied. In this case, the theoretical bound $1/T^{2}$ is represented by the $\color{blue}\textendash$ line. The number of experiments performed in both cases are the same. Note that, in the experiment shown here, for both controlled and uncontrolled cases the sensitivities scale as $1/T^{1.75\pm0.03}$ and $1/T^{0.87\pm0.02}$ respectively for the observation time T upto $80\mu$s as the de-phasing of the qubit sets in after that. The systemic lower scaling obtained in our experiment as compared to the ideal theoretical bounds stems from the incomplete preparation of the initial equal superposition state which according to eq.(\ref{eq:qfi_sat}) leads to a lower exponent of the scaling behaviour. Furthermore, the time it takes to perform the control operation is ~$30\%$ of the time for free evolution, thus leading to a lower phase accumulation than the saturated value given in eq.(\ref{eq:qfi_sat}). In order to obtain the best fit to our data, we performed least squared analysis which is discussed in the subsequent section. 

In case of amplitude estimation, as we know from eq.(\ref{eq:ampsense}), the sensitivity is independent of the observation time if no adaptive control is applied. This is evident from the measured data-points, in Figure~\ref{fig:freq_sen}(b), depicted by $\color{red}\blacksquare$ and fitted with a $\color{red}\textendash$ line with zero slope. However on application of the appropriate adaptive control the sensitivity nearly scales as $1/T$ as denoted by the $\color{blue}\bullet$ data-points and the corresponding theoretical bound shown by the $\color{blue}\textendash$ line. Thus we conclude that even the amplitude sensitivity improves upon the application of the control Hamiltonian $H_{LC}$ as compared to the sensitivity obtained without applying the control Hamiltonian. It is evident from the experimental data that in case of frequency estimation, the frequency uncertainty beats the Heisenberg scaling with observation time. In fact, an inverse $T^2$ scaling is the maximum that one can obtain theoretically for a time dependence as given in eq.~(\ref{eq:hamil}) while in the experiment we obtained $T^{-1.75}$.
%%%%%%% Figure 4
%%%%%%%       
% 
\subsection*{A proof-of-principle test case: dark matter search}
\begin{figure*}[ht] 
\centering
 \includegraphics[width=\textwidth]{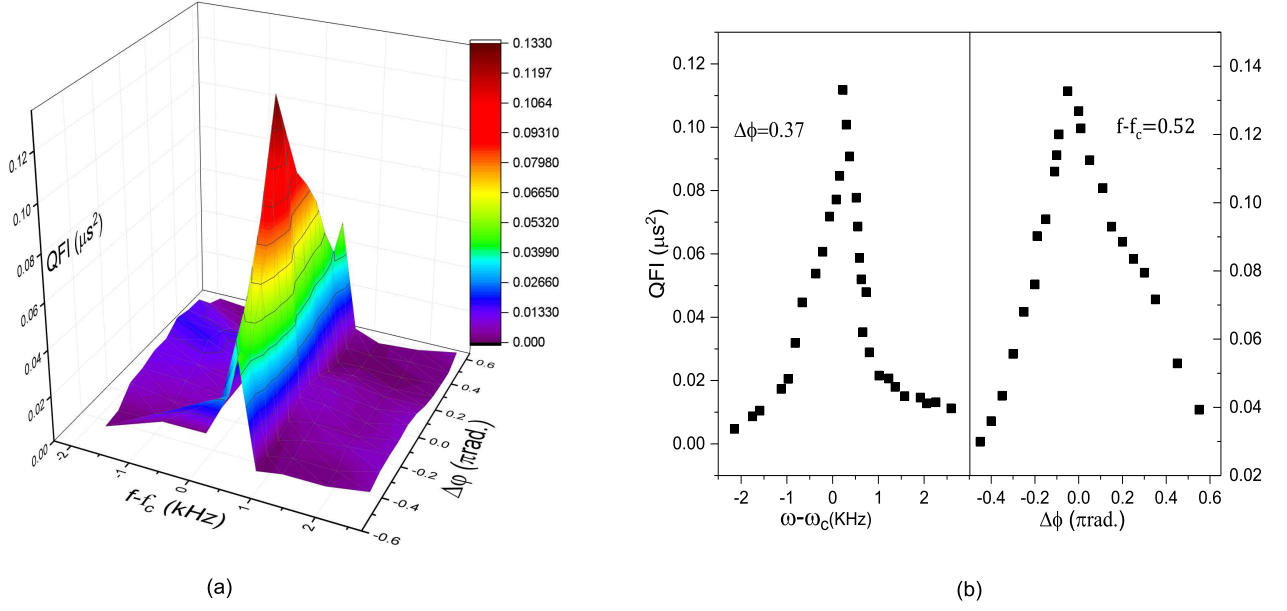} 
 \caption{(a) Experimentally measured QFI as a function of the frequency and phase of the control Hamiltonian. The estimated peak position provides the unknown frequency and phase of the external time-dependent interaction. The observation time for each point on both the plots is $75~\mu$s and $f_c = \omega_c/2\pi = 50~$kHz is the control frequency. (b) The measured QFI is plotted as a function of the frequency and phase de-tuning of the control pulses with respect to the external oscillating field. This data is extracted from the 2D surface plot in (a) close to the measured maximum of the QFI.} 
 \label{fig:3d} 
\end{figure*}

The method developed here applies to measuring the noise frequency components and their corresponding amplitudes that couple to any two level system with high precision. For sensing application, most often it is the sensitivity of the noise amplitude rather than frequency that plays an important role. A few examples of such noise are the line frequency noise appearing as time-dependent magnetic field noise, noise due to voltage fluctuation at the position of the qubit etc. These technical noises are important to the specific experimental setup. Here we show one example where a precise knowledge of the noise frequency may play a fundamental role in understanding the underlying physics independent of the experimental setup~\cite{Conrad2017,Aprile2017}. In ref.~\cite{surjeet2013}, light mass axion-like particles~(ALP is a possible dark matter candidate) have been predicted to be coupled to an electron spin by an interaction of the form (eq.~(20) of the ref.)
\begin{equation}
H_e = g_{aee}\vec{\nabla}a\cdot\vec{\sigma_e} \approx g_{aee}m_aa_0\cos(m_at)\vec{v}\cdot\vec{\sigma_e},\label{eq:DMHamil}
\end{equation}  

where $g_{aee}$ is the unknown coupling strength, $m_a$ is the unknown ALP mass, $\vec{v}\approx 3\times10^{-9}$s$^{-1}$ is the relative velocity of the ALP (also known as the ALP \enquote{wind}) with respect to the earth and $\vec{\sigma_e}$ denotes the axial component of the electron spin. The interaction originates by considering the ALP \enquote{wind} producing a free scalar field with low momentum which is oscillating in its potential as $a = a_0\cos m_at$, where $a_0$ is the amplitude of the field. The size of the actual perturbation (the time-independent part of the Hamiltonian in eq.~(\ref{eq:DMHamil})) can be estimated considering that the local dark matter density $\rho_{DM}$ is equal to the energy density of the field $\frac{1}{2} m_a^2a_0^2$ and substituting in eq.~(\ref{eq:DMHamil}), we get 
\begin{eqnarray}
\Delta E &\sim & g_{aee}v\sqrt{\rho_{DM}} \nonumber\\
&\sim & 3\times 10^{-9} s^{-1}\Big(\frac{g_{aee}}{10^{-9}GeV^{-1}}\Big)\Big(\sqrt{\frac{\rho_{DM}}{0.3\frac{GeV}{cm^3}}}\Big),\label{eq:DMsense}
\end{eqnarray}
  
where one can assume that the local dark matter density is generated by the ALPs alone ($\rho_{DM}=0.3\frac{GeV}{cm^3}$) and thus the last term becomes unity. The range $10^{-4}\ge g_{aee} \ge 10^{-10}$Gev$^{-1}$ has been ruled out by astrophysical constraints (white dwarf cooling speed)~\cite{Raffelt2008}. As conjectured, this unknown time varying ALP dark matter field couples to the spin of the valence electron of a barium ion similar to the Hamiltonian as described in eq.(\ref{eq:DMHamil}). Both the mass as well as the coupling constant are unknown parameters to be estimated. The mass corresponds to the frequency estimation in our scheme while the coupling strength corresponds to the amplitude estimation. In order to detect the existence of ALPs, we need to first estimate the sensitivity of our scheme with respect to the amplitude estimation. The amplitude sensitivity measurement provides a limit on the coupling strength. However this requires us to know the mass of the ALP which we have assumed to be $50~$kHz such that it falls somewhere in the middle of the expected range of masses for ALPs~\cite{surjeet2013}. Comparing the amplitude sensitivity with the phase uncertainty for an optimal measurement time we obtain the minimum amplitude of modulation that can barely be detected in this single ion experiment as
\begin{eqnarray}
\delta \Omega_d &=& \frac{\delta\phi}{\partial\phi/\partial \Omega_d}\nonumber\\
&\propto& \frac{1}{T\sqrt{n}},\label{eq:DMsearch}
\end{eqnarray}  

where we have used the fact that the phase noise is limited by quantum projection shot noise as shown in the Figure (S2) in the Supplementary Information. Therefore for $T=80 \mu$s measurement time and $100$ measurements, the minimal detectable frequency modulation amplitude is $\sim 1.25~$kHz at a modulation frequency of $50~$kHz. Now considering the mass of the light ALP to be $50~$kHz, by substituting eq~(\ref{eq:DMsearch}) into eq.~(\ref{eq:DMsense}), we obtain the limit on the dark matter coupling $g_{aee}$ to the electron spin in this single ion experiment as $400~$GeV$^{-1}$. Despite the bound being weak as compared to other astrophysical limits estimating $g_{aee}\le10^{-10}~$GeV$^{-1}$, it is the direct test of the limit. At the limit of our amplitude sensitivity we estimated the frequency of modulation which corresponds to the sensitivity to mass estimation of the ALPs. The frequency sensitivity of the measurement depends on two independent parameters, frequency and phase. Thus by scanning these two parameters we obtain the QFI in the neighbourhood of the target frequency and we observe a clear peak at the applied $f_c = 50~$kHz intensity modulation frequency as shown in Figure~\ref{fig:3d}(a). The QFI shows significantly high value when the OLCH frequency and phase both match to that of the test noise.

\begin{figure}[htbp]
        \centering
        \includegraphics[width=\textwidth]
        {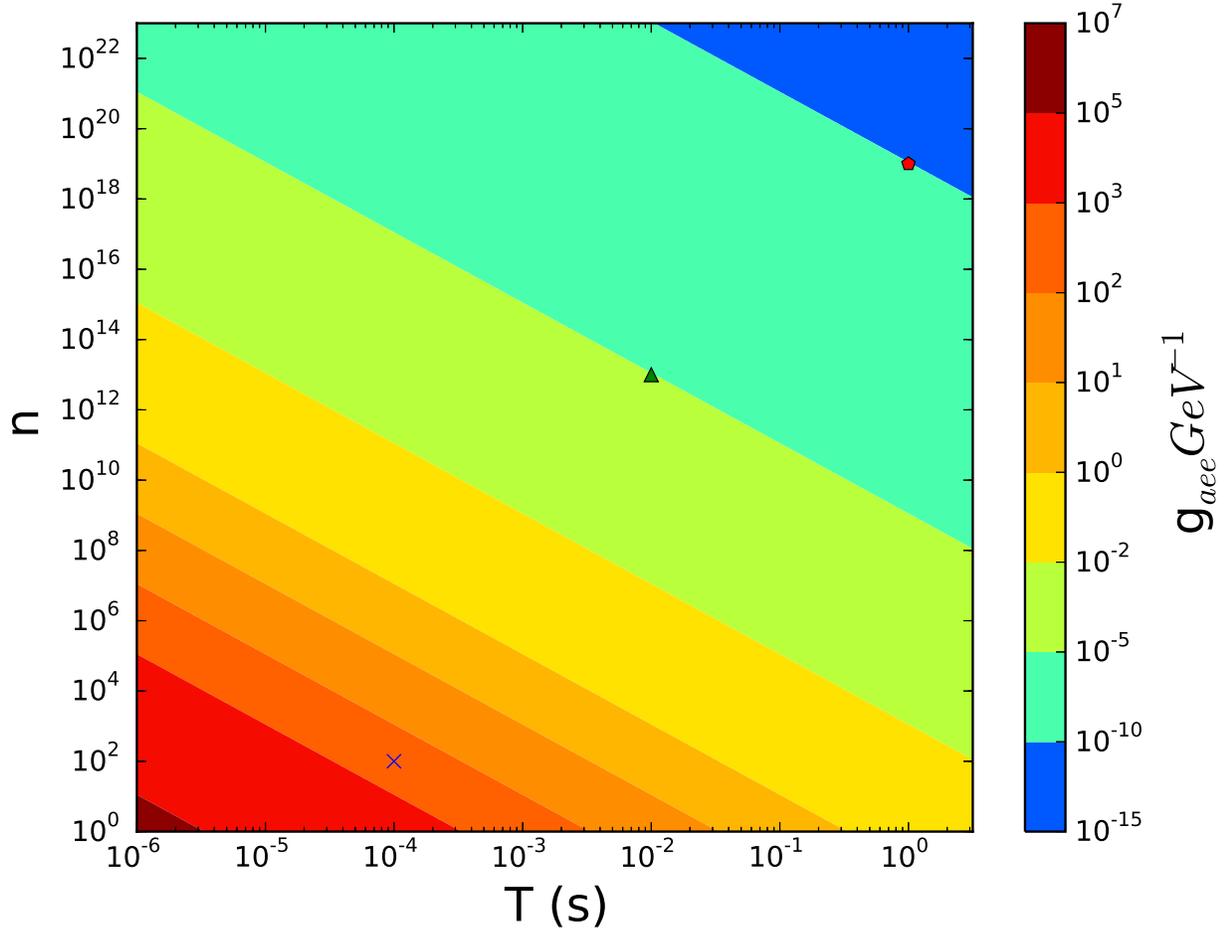}
%                \rule{10em}{0.3pt}
    \caption[Prospective experiment]{ALP dark matter coupling strength $g_{aee}$ is plotted in terms of observation time $T$ and the number of atoms or number of experiments $n$. The cross is the limit obtained from single ion experiment performed over $100$ measurements, the triangle what is achievable with gas cells, while the polygon is where the search is beyond the astrophysical range of observation}.
    \label{fig:DMpredic}
\end{figure}

It is certainly possible to improve the limit by increasing the number of independent probes $n$ as well as the measurement time as is evident from eq.~(\ref{eq:DMsearch}). However, the scheme lacks the possibility to search over a large range of ALP masses. The lock-in protocol in ref~\cite{shlomi2011} in combination with our scheme would be able to solve the problem of a large frequency range search. Once the frequency is roughly known, our adaptive measurement scheme allows precise measurement of the frequency and hence the mass. Figure~\ref{fig:DMpredic} shows the ALP dark matter coupling strength to electron axial moment as a function of the observation time $T$ and the number of probes or the single probe repeating $n$ independent measurements. The present experiment result (cross in the Figure~\ref{fig:DMpredic})  can be readily extended to atoms in a buffer gas cell (estimated limit denoted by the triangle in the Figure~\ref{fig:DMpredic}) like the Rb clocks which can probe coupling strengths of $10^{-6} - 10^{-10}~$GeV$^{-1}$ within only few measurement~\cite{Budker2016,Smullin2009}. Further improvements on these experiments can push the direct measurement limits towards the red dot where astrophysical constraints are not a binding. One of the challenges would be coherently manipulate all the spins, however it is routinely performed for magnetometry. Thus having a dedicated setup with an optimal volume and pressure of a gas cell, longer coherence time and a large number of measurements, it is possible to explore a new range of possible dark matter coupling strengths. This is particularly important since there is no other alternative method of accessing the electron-ALP coupling strength~\cite{surjeet2013}. 
 %%%%% Figure 5
%%%%
\section{Discussions}
\label{diss}
\begin{figure*}[hhh] 
\centering
\includegraphics[width=\textwidth]{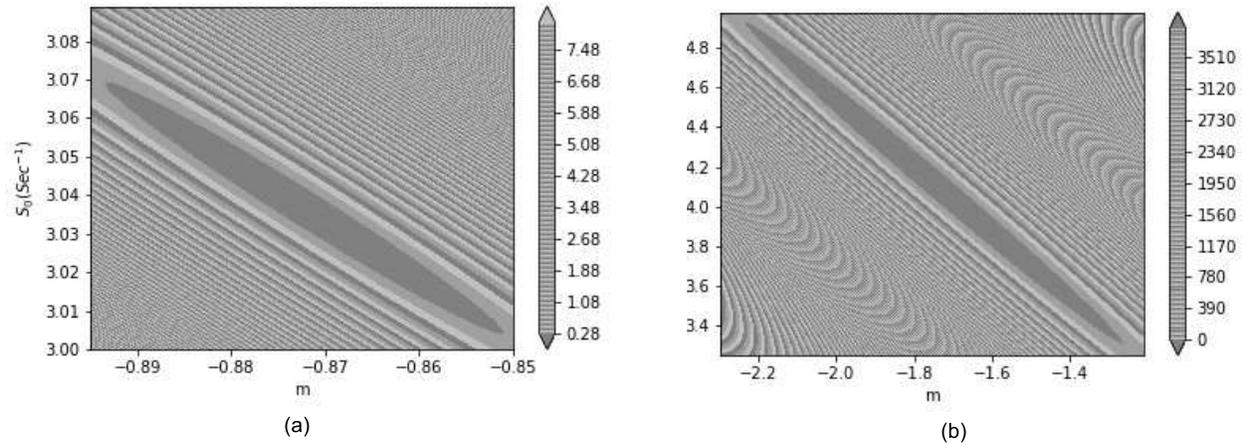} 
  
\caption{Reduced chi-squared value of the experimental data (taken upto $80\mu$s) as a function of the exponent and intercept of a linear model describing the frequency sensitivity as shown in Figure~\ref{fig:freq_sen}(a) for (a) uncontrolled measurement and (b) controlled measurement. The central ellipse denotes the $95\%$ confidence interval. In case of the controlled measurement, the ideal exponent of $-2$ lies in the $90\%$ confidence limit. The colourmaps comprise of three gray shade repeating themselves for higher values.}   
 \label{fig:chisqr}
\end{figure*}

The two main results of our experiment are shown in Figures~\ref{fig:freq_sen}(a) and \ref{fig:freq_sen}(b). According to the theory presented in ref.~\cite{Pang2017}, the limit of the achievable frequency precision with time for a time-dependent Hamiltonian can be at most $1/T^2$. In order to attain the limit, it is necessary to apply the control Hamiltonian as in eq.~(\ref{eq:olch}). The plots  in Figure~\ref{fig:freq_sen}(a) contain ideal $1/T$ and $1/T^2$ curves shown as red and blue lines for uncontrolled and controlled measurements respectively. On a closer look, the data points upto the decoherence time do not fully describe the ideal curves. Therefore to know more about the deviations, we have performed linear fit 
\begin{equation}
S = S_0 + m L_T,
\end{equation}

where $\log_{10}\frac{\partial f}{\partial \phi} = S$ and $L_T = \log_{10} T$ while $S_0$ (intercept) and $m$ (slope) are two free parameters. For each set of slope and intercept we obtained the reduced chi-square value in the vicinity of the optimal parameters which are plotted for the frequency measurement data in Figures~\ref{fig:chisqr}(a) and (\ref{fig:chisqr})(b) representing uncontrolled and controlled measurements respectively. The slope denotes the exponent of the time dependence which can at most be $-2$ for controlled measurements under ideal conditions. From this analysis, the optimal value of the exponents are $-0.87\pm 0.02$ for the uncontrolled measurement and $-1.75\pm 0.03$ for the controlled measurement. The non-ideal exponent can be explained by taking into account the error in the preparation of the equal initial superposition state and the error accumulated due to the finite time of the control pulses. It is possible to achieve the theoretical bound by improving on the initial state preparation fidelity and shortening the Rabi $\pi$-time. This requires our experimental cycle to be phase synchronized to the line frequency and the qubit laser to be intensity stabilized which are presently beyond the scope of this work.

In conclusion, we demonstrate that using quantum coherent control protocols it is possible to surpass the Heisenberg scaling on the uncertainty of a parameter estimation of a time-dependent Hamiltonian in a single ion experiment. As a proof-of-principle experiment with a single ion probe $1/T^{1.75}$ scaling upto $\sim 80~\mu$s is presented, thus extending the accessible range of noise frequencies to kHz. We further demonstrate that this technique is potentially suitable for searching light mass ALP dark matter in the mass range kHz to GHz. With the single ion probe we directly estimate the coupling strength of $50~$kHz mass ALP interacting with an electron to be lower than $400~$GeV$^{-1}$. We further propose to implement this technique on a buffer gas cell to attain sensitivity that can search new dark matter candidates using atomic probes. 

\section{Methods}
Since an atomic valence electron couples efficiently to any external magnetic or electric field, it is possible to employ either a time varying laser interaction (intensity or frequency modulation) or magnetic field (amplitude modulation) interaction to generate a time-dependent Hamiltonian or a noise. The probe is a $^{138}$Ba$^+$ which has one valence electron with half integer spin. The electronic ground state $\ket{S_{1/2},m=-1/2}$ and a meta-stable state $\ket{D_{5/2},m=-1/2}$ are used as a two level quantum system or a qubit. The probe is prepared in a coherent superposition of the qubit by a resonant narrow linewidth ($< 100$~Hz) laser phase locked to a ultra-stable optical cavity. The same laser frequency detuned from resonance and intensity modulated at a known frequency, leads to the generation of a time-dependent Hamiltonian
\begin{equation}
H_0 = -\hbar\Omega_0\sigma_z + \hbar\Omega_d\sin(\omega t)\sigma_z,\label{eq:hamil}
\end{equation}   
where the applied external off-resonant EM field leads to a constant AC Stark shift of $\hbar\Omega_0$ and a modulating energy shift of $\hbar\Omega_d$ with a modulation frequency $\omega$ as depicted in the Figure~\ref{fig:theoScheme} inset (ii). Here, $\sigma_i$ denotes $i$th Pauli spin operator. The ion is laser cooled such that its external oscillation amplitude is smaller than the wavelength of the probe laser which is also known as the Lamb-Dicke regime. This ensures that the qubit state remains mostly independent of the external motion of the ion as it is coherently manipulated. The parameters which we would like to estimate are the frequency of modulation $\omega$ and the depth of modulation $\Omega_d$. In this proof-of-principle experiment both the $\omega$ and $\Omega_d$ are set but in a sensing application, these are the unknown parameters to be measured. The total AC Stark shift including the modulation can be expressed in terms of the applied external field intensity at the ion position as ~\cite{metcalf}
\begin{equation}
\Omega_0+\Omega_d \sin \omega t = Q I(t)/ \Delta
\end{equation}
where $Q$ is a constant that depends on atomic properties of the qubit and $I(t)$ is the time-dependent intensity of the off-resonant laser which is de-tuned by a constant frequency $\Delta$ from resonance. In order to introduce a time dependence, the applied intensity of the laser is modulated about a constant value $I_0$ with an amplitude $I_d$ ($I_d\ll I_0 \implies \Omega_d\ll \Omega_0 $ in our experiment) and frequency $\omega$ as
\begin{equation}
I(t) = I_0+I_d\sin \omega t.
\label{eq:8}
\end{equation}
In the following, we show that by implementing the optimal control Hamiltonian, it is possible to estimate the $\omega$ and $\Omega_d$ with a precision that at the best scales as $1/T^2$ and $1/T$ respectively.
%%%%%%%
%
\section{Data Availability}
The data that support the findings of this study are available from the corresponding author upon reasonable request.
\section{Code Availability}
The code that is used to produce the Fig.~(\ref{fig:theoQFI}) is available from the corresponding author upon reasonable request.
\section{Acknowledgements}
This work is supported by Singapore Ministry of Education Academic
Research Fund Tier 2 (Grant No. MOE2016-T2-2-120) and NRF-CRP (Grant No. NRF-CRP14-2014-02). MM would like to thank Dr. S. Rajendran for many useful discussion while TD would like to thank Dr. D. Yum for his help in the experiment.
\section{Competing Interests}
The authors declare that there are no competing interests.
\section{Author Contribution}
TD has contributed towards setting up the experiment and data taking while MM has developed the concept, numerical analysis and applicability of the protocol to other fields. Both the authors contributed equally towards analysing the data, writing the manuscript and presenting the final conclusions. 
%%%%%%%%%%%%%%%%%%%%%%%%

%%%%%%%%%% FIGURE LEGENDS
%%%%%%%%%%%% ALL FIGURES%%%%%%%%%%%%%%
%%FIGURE 1
\section{Figure Legends}
%
%%%%%%%FIGURE 2
%%%%%%FIGURE 3
%%%%%%%%%FIGURE 4
%%%%%%%%%%FIGURE 5
%
%%%%%%%%%%%%%
\end{document}